\documentclass[pre,superscriptaddress,twocolumn,showpacs]{revtex4}
\usepackage{epsfig}
\usepackage{latexsym}
\usepackage{amsmath}
\begin {document}
\title {Molecular dynamics simulations of ballistic annihilation}
\author{Adam Lipowski}
\affiliation{Faculty of Physics, Adam Mickiewicz University, 61-614
Pozna\'{n}, Poland}
\author{Dorota Lipowska}
\affiliation{Institute of Linguistics, Adam Mickiewicz University,
60-371 Pozna\'{n}, Poland}
\author{Ant\'onio L.~Ferreira}
\affiliation{Department of Physics, University of Aveiro, 3810
Aveiro, Portugal}
\begin {abstract}
Using event-driven molecular dynamics we study one- and
two-dimensional ballistic annihilation. We estimate exponents
$\xi$ and $\gamma$ that describe the long-time decay of the number
of particles ($n(t)\sim t^{-\xi}$) and of their typical velocity
($v(t)\sim t^{-\gamma}$). To a good accuracy our results confirm
the scaling relation $\xi + \gamma =1$. In the two-dimensional
case our results are in a good agreement with those obtained from
the Boltzmann kinetic theory.
\end{abstract}
\maketitle
Kinetics of reacting particle systems is recently intensively
studied. Such systems model a wealth of different phenomena and in
addition offer an important testing ground for nonequilibrium
statistical mechanics. In lattice formulation such systems are
much easier to analyze and in some cases even exact results could
be obtained. More realistic, off-lattice (ballistic) versions are
at the same time much more difficult to examine, but even here
some analytical insight is possible~\cite{LEYVRAZ}.

Despite being one of the simplest reacting particle systems,
single-species ballistic annihilation, $A+A\rightarrow 0$, so far
evaded exact solution. Consequently, our understanding of the
kinetics of this process comes mainly from approximate methods. To
describe ballistic annihilation one usually calculates density of
particles $n(t)$ and their typical velocity $v(t)$ defined using
the second moment of the time-dependent velocity distribution
$f(\boldsymbol{v},t)$
\begin{equation}
v(t)^2=\frac{1}{n(t)}\int v^2f(\boldsymbol{v},t)d\boldsymbol{v}
\label{eq1}
\end{equation}
Of particular interest are then the exponents $\xi$ and $\gamma$
that describe the asymptotic decay of these quantities
\begin{equation}
n(t)\sim t^{-\xi},\ \ v(t)\sim t^{-\gamma} \label{eq2}
\end{equation}
A simple power counting or scaling arguments can be used to show
that
\begin{equation}
\xi + \gamma = 1 \label{eq3}
\end{equation}
Although a rigorous justification of this relation is missing,
Piasecki et al.~{\cite{DROZ}} analysing BBGKY-like hierarchy
presented convincing arguments supporting relation~(\ref{eq3}).

A lot of efforts was made to calculate $\xi$ and
$\gamma$~\cite{REDNER,TRIZAC,TRIZKRAP,DROZ,CHOPARD,KRAPSIRE}.
However, it would be desirable to obtain more accurate estimations
of these exponents since it could increase our understanding in
this field. For example, some arguments suggest~\cite{DROZ} that
in dimensions $d>1$ Boltzmann kinetic theory provides an accurate
description of such a process. However, a direct verification of
this hypothesis, using molecular dynamics simulations, is not yet
fully completed mainly due to too small number of particles that
were taken into account. Let us also notice that current
estimations of $\xi$ for $d=1,2$ and 3 are surprisingly close to
the result obtained in the so-called reaction-controlled limit
($\xi=4d/(4d+1)$)~\cite{TRIZKRAP}. Further check whether this
exponent could be given as such a simple fraction would be
desirable. Let us notice that a closely related process, namely
ballistic aggregation, is solvable in $d=1$ and some of its
exponents are indeed simple fractions~\cite{POMEAU}.

In the present paper we describe results of extensive molecular
dynamics simulations of ballistic annihilation in $d=1$ and 2.
Since in such a process particles interact (i.e., annihilate) only
upon collision efficient simulations can be made with the use of
the so-called event-driven dynamics~\cite{RAPAPORT}.  One of the
difficulties in this technique is the search for the time of the
nearest collision. Most efficient algorithms to locate such an
event arrange data on the heap tree~\cite{KNUTH}. For event-driven
simulations for $d>1$ systems it is also essential to include
sectorization and search for a collision partner only within a
given sector and its nearest neighbours. Heap tree searching and
sectorization were already used in event-driven dynamics of
various hard-core or granular systems~\cite{BORIS,CORDERO,LUDING}
but to our knowledge not in the ballistic annihilation. Since it
substantially improves the numerical performance, our method
implements these techniques. The fact that at the collision
particles annihilate, simplifies the algorithm comparing to the
non-annihilating systems. This is because there is no need to
examine post-collision events of colliding particles. In the
following we discuss obtained numerical results.

In the $d=1$ case we place $N$ point-like particles on a line
interval of length $L=N$. Such a choice keeps the initial density
of particles $n=N/L$ constant (and equal to unity) and that will
allow us the comparison of results for different values of $N$.
Initially velocities have a distribution $f(\boldsymbol{v},t=0)$
that either will be gaussian or  it will have a
$\boldsymbol{v}=\boldsymbol{0}$ singularity
$f(\boldsymbol{v},t=0)\sim |\boldsymbol{v}|^{-\mu}$ with a
characteristic exponent $\mu$. Periodic boundary conditions are
used. We measured the density of particles $n(t)$ at time $t$ and
their typical velocity $v(t)$ defined using Eq.~(\ref{eq1}). To
estimate the exponents $\xi$ and $\gamma$ the system must reach
the asymptotic regime. However, for finite $N$ the long-time
behaviour is affected by poor statistics as well as by finite-size
effects and that hinders precise determination of $\xi$ and
$\gamma$. To take these effects into account we calculated
time-dependent exponents $\xi(t)$ and $\gamma(t)$ using results
(i.e., $n(t)$ and $v(t)$) spreading over one decade around a given
value of $t$.

\begin{figure}[!t]
\centerline{ \epsfxsize=9cm \epsfbox{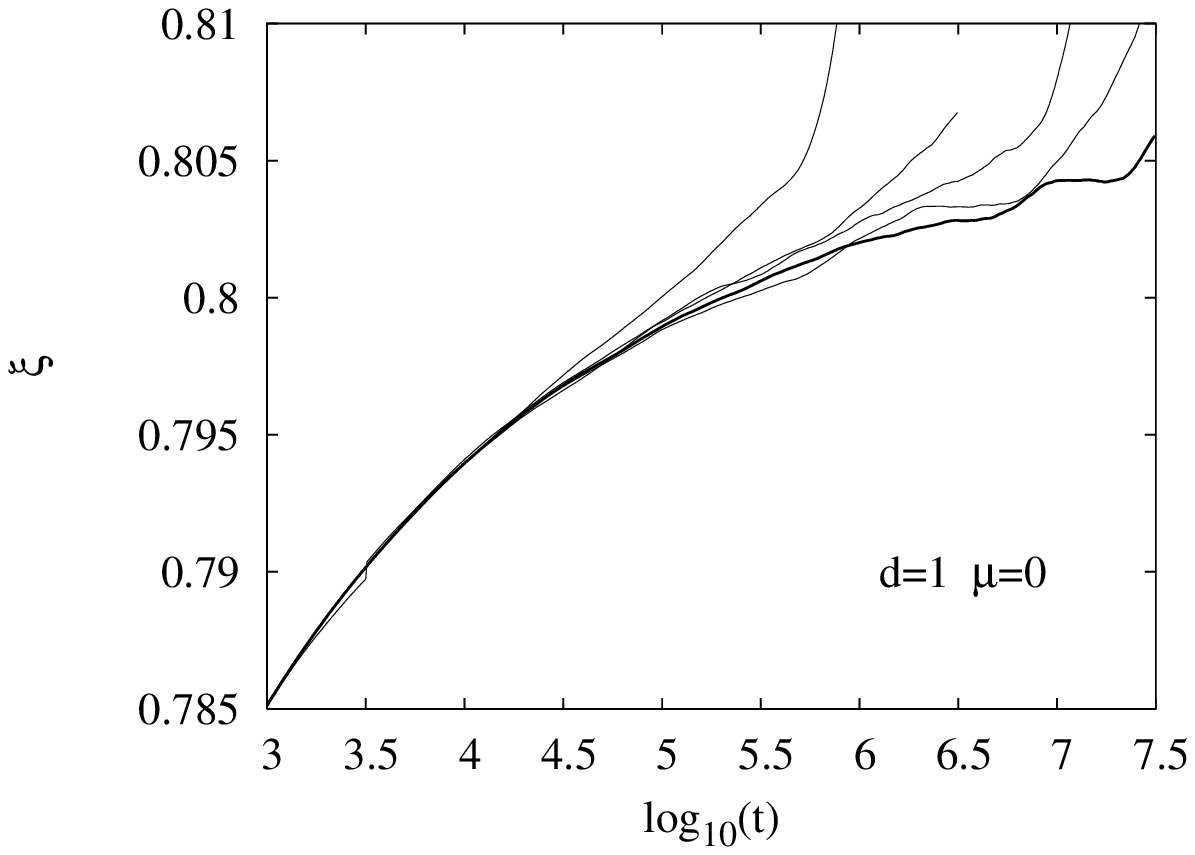} }
\caption{The exponent $\xi$ as a function of $t$ for $N=10^6,\
5\cdot10^6,10^7, 2\cdot 10^7$, and $3\cdot 10^7$ (thick line).
Initially, velocities have a gaussian distribution ($\mu=0$).}
\label{ksigd1m0}
\end{figure}

\begin{figure}[!ht]
\centerline{ \epsfxsize=9cm \epsfbox{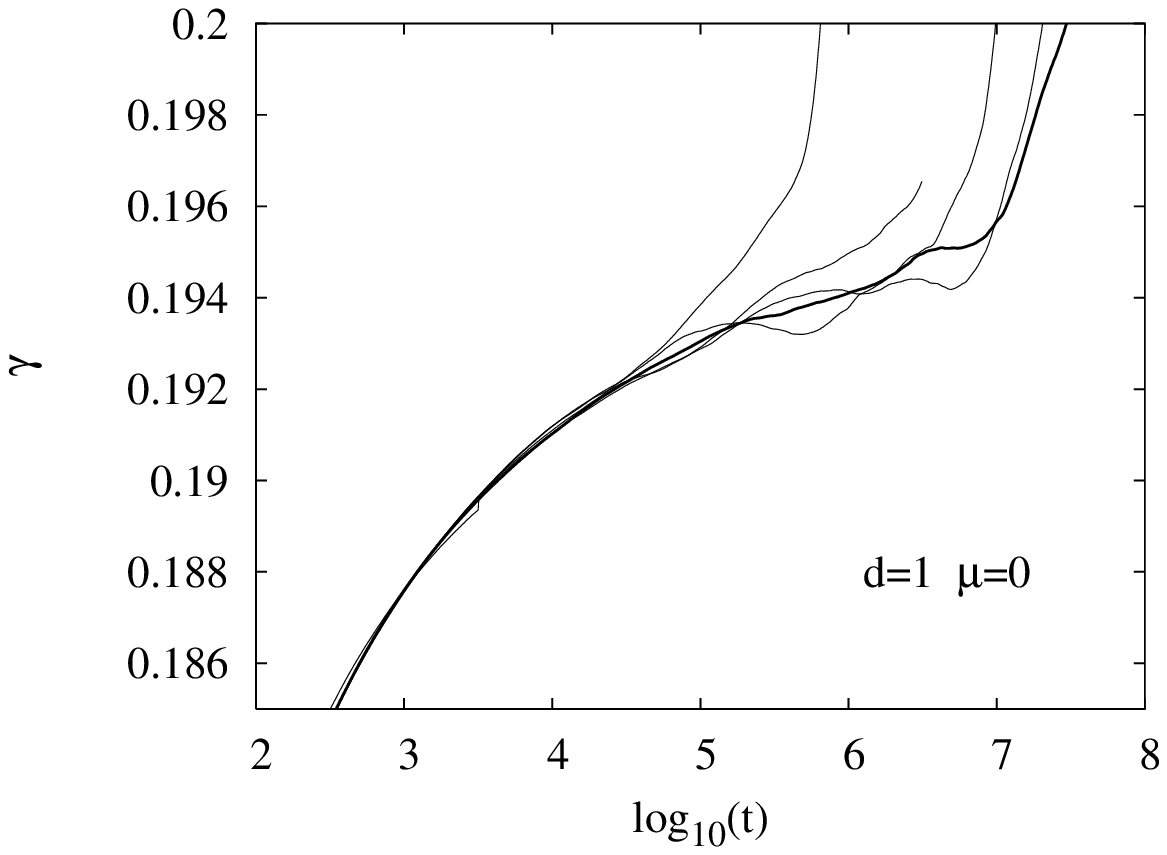} }
\caption{The exponent $\gamma$ as a function of $t$ for $N=10^6,\
5\cdot 10^6,10^7, 2\cdot 10^7$, and $3\cdot 10^7$ (thick line).
Initially, velocities have a gaussian distribution ($\mu=0$). }
\label{gamgd1m0}
\end{figure}

For the gaussian initial velocity distribution our results are
shown in Fig.~\ref{ksigd1m0}-\ref{gamgd1m0}. Presented values are
averages over $3\cdot 10^4$ (for $N=3\cdot 10^7$) to $10^5$ (for
$N=10^6$) samples. Although examined systems were rather large
($N_{max}=3\cdot 10^7$), it is still difficult to make precise
evaluation of exponents. Our results, $\xi=0.805(2)$ and
$\gamma=0.195(2)$, satisfy Eq.~(\ref{eq3}), but we admit that the
estimation of errors is based mainly on the visual inspection of
data and the hope that further increase of the system size will
not change much the final estimations. Although the difference is
very small, such a value of $\xi$ seems to exclude the possibility
$\xi=0.8$ obtained (exactly) for the $d=1$ annihilation process in
the reaction-controlled limit. Our estimation of $\xi$ also
slightly differs from previous molecular dynamics simulations
result 0.785(5) made by Rey {\it et al.}~\cite{DROZ1998} albeit
with much smaller number of particles (they simulated systems with
up to $262144$ particles). Let us notice, that computing time in
the search algorithm of Rey {\it et al.} increases as
$O(N^{5/4}{\rm ln}(N))$ while for the heap tree search (that we
used) it increases only as $O(N{\rm ln}(N))$. Various computations
based on the solution of the Boltzmann equation, and thus based on
the molecular chaos hypothesis, predicts $\xi$ close to
$0.77$~\cite{KRAPSIRE,TRIZAC}. Disagreement between this result
and our estimation shows that in $d=1$ Boltzmann kinetic theory
has only limited validity.
\begin{figure}[!ht]
\centerline{ \epsfxsize=9cm \epsfbox{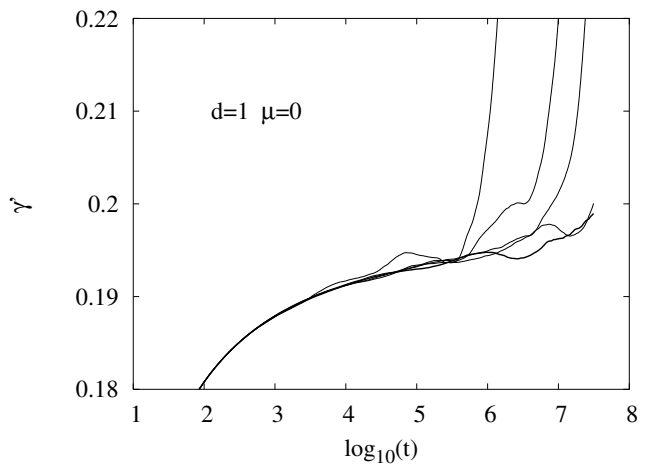} }
\caption{The time dependence of the exponent $\gamma'$ that
describes the decay of the typical velocity defined via the fourth
moment. Calculations were made for $N=10^6,\ 5\cdot 10^6,10^7,
2\cdot 10^7$, and $3\cdot 10^7$ (thick line). Initially,
velocities have a gaussian distribution ($\mu=0$). }
\label{v4d1m0}
\end{figure}
\begin{figure}[!ht]
\centerline{ \epsfxsize=9cm \epsfbox{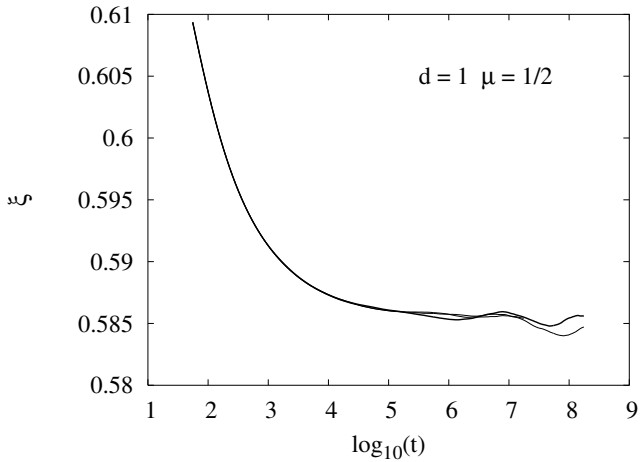} }
\caption{The exponent $\xi$ as a function of $t$ for $N=10^6,\
3\cdot10^6$, and $10^7$ (thick line). Initially, velocities have a
distribution with $\mu=1/2$ singularity.} \label{ksid1m12}
\end{figure}

To check for a possible multi-scaling we calculated the typical
velocity $v'(t)$ using the fourth-moment analogue of
Eq.~(\ref{eq1}). Fig.~\ref{v4d1m0} shows the exponent $\gamma'$
that describes the decay of $v'(t)$. Within our numerical
precision $\gamma=\gamma'$ and thus $v(t)$ as defined in
Eq.~(\ref{eq1}) is the only characteristic velocity in this
problem.

\begin{figure}
\centerline{ \epsfxsize=9cm \epsfbox{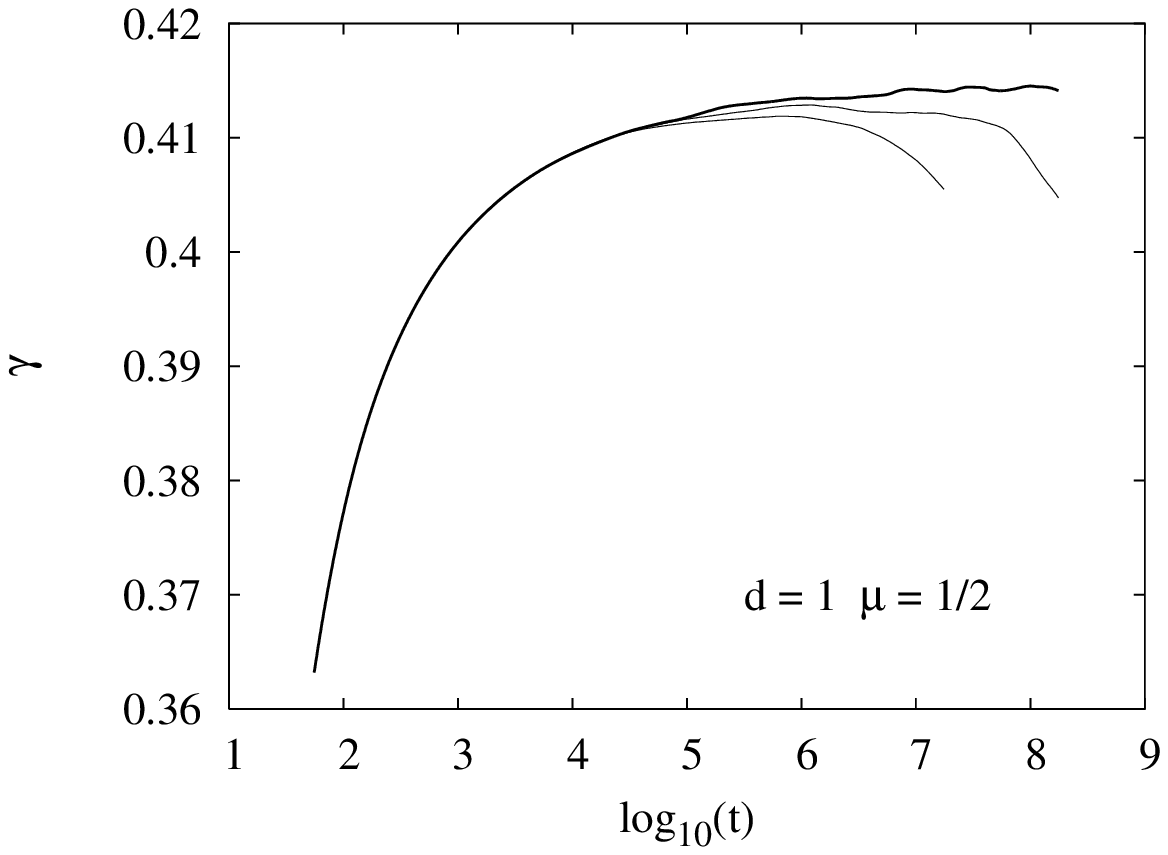} }
\caption{The exponent $\gamma$ as a function of $t$ for $N=10^6,\
3\cdot10^6$, and $10^7$ (thick line). Initially, velocities have a
distribution with $\mu=1/2$ singularity.} \label{gamd1m12}
\end{figure}

We also performed calculations for the singular initial velocity
distribution with $\mu=1/2$. In this case we estimate (see
Figs.~\ref{ksid1m12}-\ref{gamd1m12}) $\xi=0.585(2)$ and
$\gamma=0.415(2)$. This result satisfies Eq.~(\ref{eq3}) and
improves over previous estimations of $\xi$ reported in the
literature that range from 0.5 to
0.6~\cite{REDNER,REDNER94,TRIZAC}.

In the one dimensional case topological constraints imply that for
a given particle potential partners for a collision are only two
of its nearest neighbours. In two dimensions this is no longer the
case and in principle any pair of particles can collide. For large
$N$ it leads to a prohibitively large number of potential
collisions that should be examined but only very few of them will
actually take place. To overcome this difficulty one can divide
available space into sectors and look for potential collisions
only within a sector or neighbouring sectors. More details on this
technique can be found elsewhere~\cite{BORIS,LUDING}.

\begin{figure}
\centerline{ \epsfxsize=9cm \epsfbox{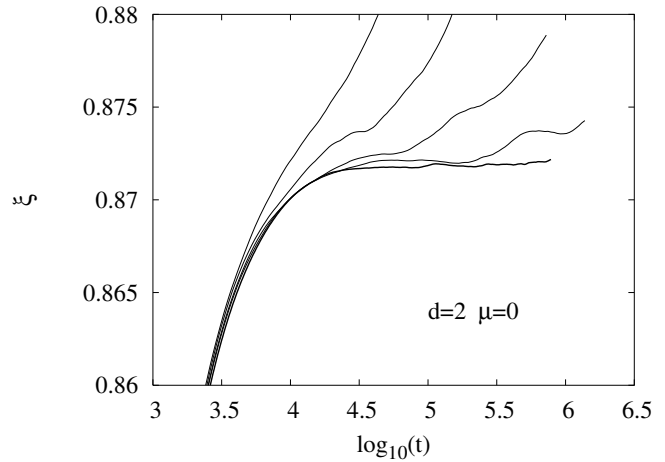} }
\caption{The exponent $\xi$ as a function of $t$ for $N=300^2,\
500^2,\ 1000^2,\ 2000^2$, and $3000^2$ (thick line). Initially,
velocities have a gaussian distribution ($\mu=0$).}
\label{ksid2m0}
\end{figure}

\begin{figure}
\centerline{ \epsfxsize=9cm \epsfbox{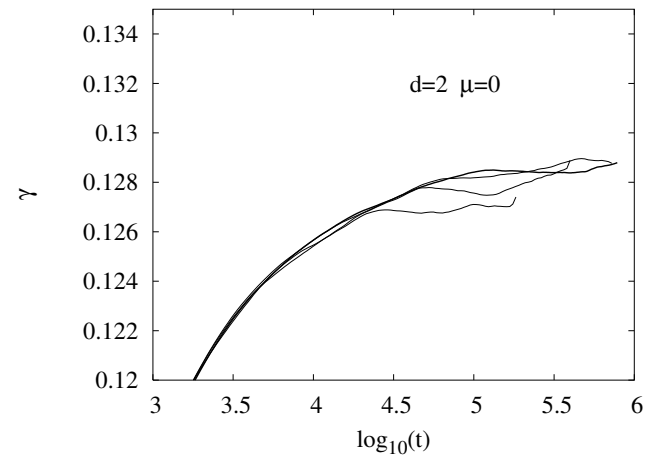} }
\caption{The exponent $\gamma$ as a function of $t$ for $N=300^2,\
500^2,\ 1000^2$, and $3000^2$ (thick line). Initially, velocities
have a gaussian distribution ($\mu=0$).} \label{gamd2m0}
\end{figure}

For the gaussian initial velocity distribution results of our
simulations are shown in Figs~\ref{ksid2m0}-\ref{gamd2m0}. In two
dimensions size of particles $r$ relative to the linear system
size $L$ becomes a relevant variable that might affect e.g., the
time needed to reach asymptotic regime. Presented results are
obtained for the packing fraction $f=\pi r^2N/L^2=0.0079$ but
similar results were obtained for $f=0.031$. After averaging over
$10^3-10^4$ samples, we estimate $\xi=0.872(2)$ and
$\gamma=0.129(2)$. Such results satisfy Eq.~(\ref{eq3})and are in
a very good agreement with calculations based on the Boltzmann
equation~\cite{TRIZAC,DROZ}, confirming thus validity of molecular
chaos hypothesis in this case. Let us notice that our estimation
of $\xi$ definitely excludes the reaction-controlled value
$\xi=8/9$. Moreover, the maximal number of particles examined in
our approach ($9\cdot 10^6$) is almost 20 times larger that for
previously reported $d=2$ event-driven simulations~\cite{DROZ}.

\begin{figure}
\centerline{ \epsfxsize=9cm \epsfbox{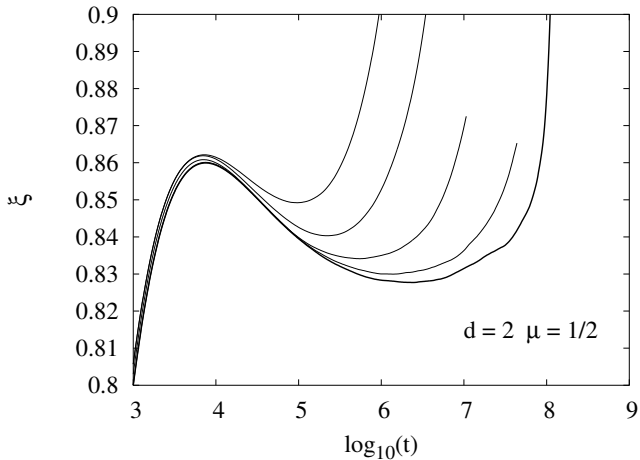} }
\caption{The exponent $\xi$ as a function of $t$ for $N=300^2,\
500^2,\ 1000^2,\ 2000^2$, and $3000^2$ (thick line). Initially,
velocities have a distribution with $\mu=1/2$ singularity.}
\label{ksid2m12}
\end{figure}

\begin{figure}
\centerline{ \epsfxsize=9cm \epsfbox{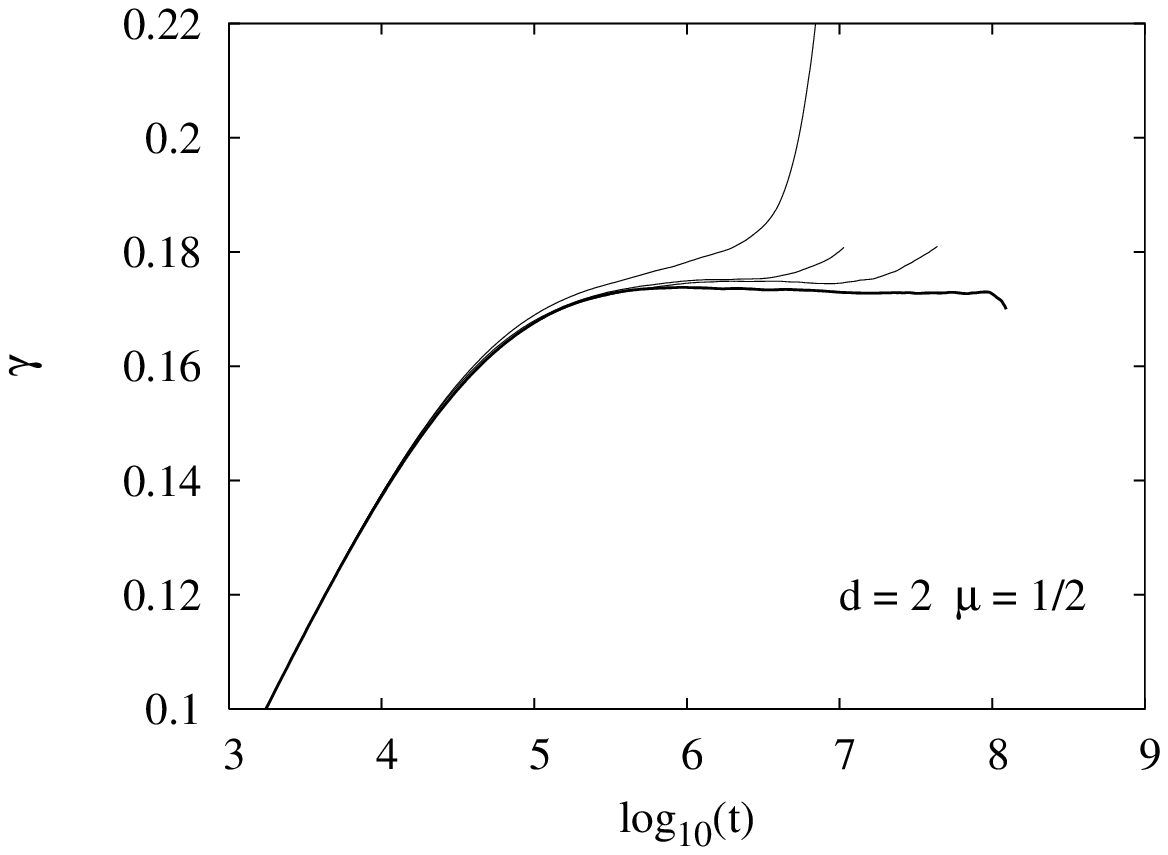} }
\caption{The exponent $\gamma$ as a function of $t$ for $N=500^2,\
1000^2,\ 2000^2$, and $3000^2$ (thick line). Initially, velocities
have a distribution with $\mu=1/2$ singularity.} \label{gamd2m12}
\end{figure}

We also made calculations for the $\mu=1/2$ case and the results
are presented in Fig.~\ref{ksid2m12}-\ref{gamd2m12}. We estimate
$\xi=0.83(1)$ and $\gamma=0.17(1)$. Similar value of $\xi$ was
recently obtained using numerical integration of the Boltzmann
equation~\cite{TRIZAC}. In Table~\ref{table1} we collect all our
final results.

\begin{table}[!ht]
\begin{tabular}{|@{\hspace{3mm}}c@{\hspace{3mm}}|@{\hspace{1mm}}c@{\hspace{1mm}}|c|c|c|c|c|}
\hline
d & $\mu$ & $\xi$ & $\gamma$ & $\xi$~\cite{REDNER94} & $\xi$~\cite{TRIZKRAP} & $\xi$~\cite{TRIZAC}\\
\hline
1 & 0 & 0.805(2) & 0.195(2) & 0.666 & 0.8 & 0.769\\
\cline{2-7}
& 1/2 & 0.585(2) & 0.415(2) & 0.5 & -- & 0.6\\
\hline
2 & 0 & 0.872(2) & 0.129(2) & 0.8 & 0.75 & 0.87\\
\cline{2-7}
& 1/2 & 0.83(1) &  0.17(1) & 0.75 & -- & 0.84\\
\hline
\end{tabular}
\caption{\label{table1} Exponents $\xi$ and $\gamma$ as calculated
in the present paper. Estimations of $\xi$, based on the analysis
of the Boltzmann equation~\cite{REDNER94,TRIZAC} and in the
reaction-controlled limit~\cite{TRIZKRAP} are presented in the
last three columns.}
\end{table}

In conclusion, implementing a heap-tree search algorithm and
sectorization we made extensive event-driven molecular dynamics
simulations of the ballistic annihilation. Obtained estimations of
decay exponents $\xi$ and $\gamma$ obey the scaling relation $\xi
+ \gamma = 1$ and improve over previously reported results. Our
calculations suggest that the typical velocity as defined via the
second moment sets the only characteristic scale in the problem.
Very good agreement of our simulations with predictions of
Boltzmann kinetic theories in the $d=2$ case confirms validity of
the latter approach. In the $d=1$ case small differences between
results of these two approaches exist. It would be interesting to
extend our approach to some other model phenomena whose kinetics
is still not fully understood as e.g., ballistic
aggregation~\cite{TRIZKRAP,TRIZHANS} or probabilistic ballistic
annihilation~\cite{COPPEX}.

Acknowledgments: A.~Lipowski acknowledges the research grant 1
P03B 014 27 from KBN. A.~L.~Ferreira acknowledges financial
support from
 Funda\c c\~ao Ci\^encia e Tecnologia (FCT) of Portugal.


\begin{thebibliography}{}
\bibitem{LEYVRAZ} F.~Leyvraz, Phys.~Rep.~{\bf 383}, 95 (2003).
\bibitem{REDNER} E.~Ben-Naim, S.~Redner, and F.~Leyvraz, Phys. Rev. Lett. {\bf 70}, 1890 (1993).
\bibitem{TRIZAC} E.~Trizac, Phys.~Rev.~Lett.~{\bf 88}, 160601 (2002).
\bibitem{TRIZKRAP} E.~Trizac and P.~L.~Krapivsky, Phys.~Rev.~Lett.~{\bf 91}, 218302 (2003).
\bibitem{DROZ} J.~Piasecki, E.~Trizac, and M.~Droz, Phys.~Rev.~E {\bf 66}, 066111 (2002).
\bibitem{CHOPARD} B.~Chopard, A.~Masselot, and M.~Droz, Phys. Rev. Lett. {\bf 81}, 1845 (1998).
\bibitem{KRAPSIRE} P.~L.~Krapivsky and C.~Sire, Phys.~Rev.~Lett.~{\bf 86}, 2494 (2001).
\bibitem{POMEAU} G.~F.~Carnevale, Y.~Pomeau, and W.~R.~Young, Phys.~Rev.~Lett.~{\bf 64},
2913 (1990). L.~Frachebourg, Phys.~Rev.~Lett.~{\bf 82}, 1502
(1999).
\bibitem{RAPAPORT} D.~C.~Rapaport, {\it The Art of Molecular Dynamics Simulations},
(Cambridge University Press, Cambridge, 1995).
\bibitem{KNUTH} D.~E.~Knuth, {\it The Art of Computer Programming},
(Addison-Wesley, Reading, MA, 1968).
\bibitem{BORIS} B.~D.~Lubachevsky, J.~Comput.~Phys.~{\bf 94}, 255 (1991).
\bibitem{CORDERO} M.~Mar\'in and P.~Cordero, Comput.~Phys.~Commun.~{\bf 92}, 214 (1995).
\bibitem{LUDING} S.~Miller and S.~Luding, J.~Comput.~Phys.~{\bf 193}, 306 (2004).
\bibitem{REDNER94} E.~Ben-Naim, P.~L.~Krapivsky, F.~Leyvraz, and S.~Redner, J.~Phys. Chem. {\bf 98}, 7284 (1994).
\bibitem{DROZ1998} P.-A.~Rey, M.~Droz, and J.~Piasecki, Phys.~Rev.~E {\bf 57}, 138 (1998).
\bibitem{TRIZHANS} E.~Trizac and J.-P.~Hansen, Phys.~Rev.~Lett.~{\bf 74}, 4114 (1995).
\bibitem{COPPEX} F.~Coppex, M.~Droz, and E.~Trizac, Phys.~Rev.~E {\bf 72} 021105 (2005).
\end{thebibliography}
\end {document}